\nonstopmode

\newcommand{\eV}{\U{eV}}

\newcommand{\U}[1]{\,{\rm{#1}}}
\newcommand{\I}[1]{_{\mathrm{#1}}}
\newcommand{\mul}{\cdot}

\renewcommand{\frac}[2]{{{#1}\over{#2}}}
\newcommand{\dfrac}[2]{{\displaystyle{#1}\over\displaystyle{#2}}}

\documentclass[prl,aps,twocolumn,showpacs,nopreprintnumbers,superscriptaddress]{revtex4}
\usepackage{bm,bbm,amssymb,graphicx}
\usepackage[nativepdf,bookmarks,bookmarksopen,bookmarksnumbered,raiselinks,%
pdftitle={Refraction and absorption of x rays by laser-dressed atoms},%
pdfauthor={Christian Buth, Robin Santra, Linda Young},%
pdfsubject={Atomic physics},%
pdfkeywords={quantum optics, ab initio, laser dressing, x rays, %
photoabsorption, cross section, polarization dependence, nonperturbative, %
laser-matter interaction, neon, dispersion, electromagnetically induced %
transparency, EIT, pulse shaping, index of refraction}]{hyperref}

\begin{document}
\title{Refraction and absorption of x~rays by laser-dressed atoms}
\author{Christian Buth}
\thanks{Present address: Department of Physics and Astronomy, Louisiana State
University, Baton Rouge, Louisiana~70803, USA}
\affiliation{Argonne National Laboratory, Argonne, Illinois~60439, USA}
\author{Robin Santra}
\affiliation{Argonne National Laboratory, Argonne, Illinois~60439, USA}
\affiliation{Department of Physics, University of Chicago, Chicago,
Illinois~60637, USA}
\author{Linda Young}
\affiliation{Argonne National Laboratory, Argonne, Illinois~60439, USA}
\date{10 November 2010}

\begin{abstract}
X-ray refraction and absorption by neon atoms under the influence of an
$800\, \U{nm}$~laser with an intensity of~$10^{13}\, \U{W/cm^2}$ is
investigated.
For this purpose, we use an \emph{ab initio} theory suitable for optical
strong-field problems.
Its results are interpreted in terms of a three-level model.
On the $\mathrm{Ne} \, 1s \to 3p$~resonance, we find
electromagnetically induced transparency~(EIT) for x~rays.
Our work opens novel perspectives for ultrafast x-ray pulse shaping.
\end{abstract}

%
%
%

\preprint{arXiv:0805.2619}
\maketitle

\begin{figure}
  \begin{center}
    \includegraphics[clip,width=1in]{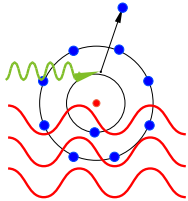}
    \caption{(Color online) Laser-dressed neon atom probed by x~rays.}
    \label{fig:Ne_ion}
  \end{center}
\end{figure}

We consider atoms in the light of a linearly polarized laser
which are probed by linearly polarized x~rays [Fig.~\ref{fig:Ne_ion}].
Such a setting is referred to as a two-color problem and has proved to be
a very beneficial to probe strong-field physics and to manipulated
x~rays~\cite{Buth:TX-07,Loh:QS-07,Buth:ET-07,Santra:SF-07,Buth:LA-08,%
Peterson:XR-08,Buth:AR-08}.

We assume a Ti:sapphire laser at $800 \U{nm}$ with $10^{13}\,$W/cm$^2$.
The atomic ground-state electrons remain essentially unperturbed in the laser light;
they are neither excited nor ionized.
However, laser dressing of the core-excited states introduces strong-field physics
because, for the laser parameters employed here,
the Keldysh parameter~\cite{Keldysh:-65} is~$\gamma = \sqrt{I_{1s^{-1}\,3p}/(2 \,
U\I{p})} = 1.5 \approx 1$.
The ionization potential of a core-excited state with the electron in the
$3p$~Rydberg orbital is denoted by~$I_{1s^{-1}\,3p}$ and
the ponderomotive potential of the laser is~$U\I{p}$~\cite{Buth:ET-07}.

For $\gamma \approx 1$, neither perturbation theory nor a tunneling description is adequate.
Therefore, we have developed an \emph{ab initio} theory described in
Ref.~\onlinecite{Buth:TX-07}.
The electronic structure of the atom is described in the Hartree-Fock-Slater
approximation where a complex absorbing potential~(CAP) is used to treat
continuum electrons.
Nonrelativistic quantum electrodynamics is used to treat the atom in the
two-color light.
The interaction with the laser is described in terms of a Floquet-type matrix
whereas the x-ray probe is treated in second-order (one-photon) perturbation
theory~\cite{Buth:NH-04}.
For core-excited states, we need to take into account the Auger decay rate of a
$K$-shell hole in neon~$\Gamma_{1s} = 0.27 \eV$.
The absorption cross section for x~rays with photon energy~$\omega\I{X}$ is
found to be
\begin{equation}
  \sigma_{1s}(\omega\I{X}, \vartheta\I{LX}) =
      \sigma_{1s}^{\parallel}(\omega\I{X}) \, \cos^2 \vartheta\I{LX}
    + \sigma_{1s}^{\perp}    (\omega\I{X}) \, \sin^2 \vartheta\I{LX} \; .
\end{equation}
where $\vartheta\I{LX}$ denotes the angle between the laser and x-ray
polarization vectors~\cite{Buth:TX-07}.
The cross section for parallel x-ray and laser polarization vectors
is given by~$\sigma_{1s}^{\parallel}(\omega\I{X})$;
it is~$\sigma_{1s}^{\perp}(\omega\I{X})$ for perpendicular vectors.
The calculations were carried out with the \textsc{dreyd} and
\textsc{pulseprop} programs~\cite{fella:pgm-V1.3.0}.
We use the computational parameters from
Refs.~\onlinecite{Buth:TX-07,Buth:ET-07}.
For the index of refraction [Fig.~\ref{fig:NeRI}], we need 400~radial
wavefunctions.

\begin{figure}[bht]
  \begin{center}
    \includegraphics[clip,width=\hsize]{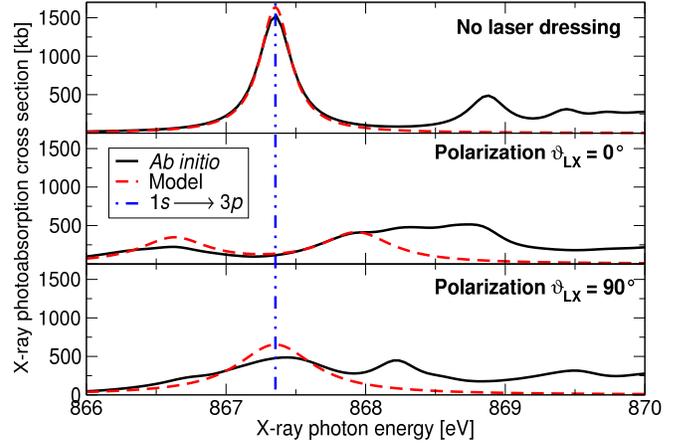}
    \caption{(Color online) X-ray photoabsorption cross section of neon near the
             $K$~edge with laser dressing and without it.}
    \label{fig:NeI_model}
  \end{center}
\end{figure}

The calculated x-ray photoabsorption cross section of a neon atom is displayed in
Fig.~\ref{fig:NeI_model} with laser dressing and without.
The prominent absorption feature in the top panel at~$867.4 \eV$ arises due to
the $1s \to 3p$~resonance.
The weaker peaks are associated with $1s \to np$ transitions with $n \ge 4$.
The $K$~edge lies at $870.2 \eV$.
The most eye-catching impact of the laser dressing can be seen in the
vicinity of the $1s \to 3p$ resonance, which becomes transparent
for~$\vartheta\I{LX} = 0^{\circ}$~\cite{Buth:ET-07}.

\begin{figure}
  \begin{center}
    \includegraphics[clip,width=2in]{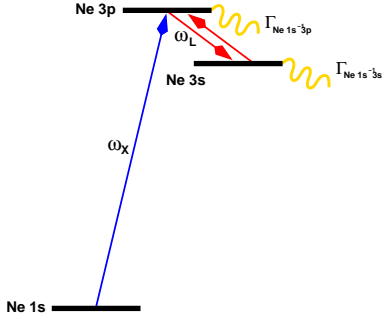}
    \caption{(Color online) $\Lambda$-type three-level model for neon.}
    \label{fig:three-level}
  \end{center}
\end{figure}

To understand the influence of the laser, we make a $\Lambda$-type model for
neon shown in Fig.~\ref{fig:three-level}.
We use the states~$1s^{-1} \, 3p$, $1s^{-1} \, 3s$, and the atomic ground
state.
The laser photon energy is~$\omega\I{L}$ whereas $\Gamma_{\mathrm{Ne} \,
1s^{-1} \, 3p}$ and $\Gamma_{\mathrm{Ne} \, 1s^{-1} \, 3s}$ denote the decay
widths of the respective core-excited states.
We find the overall agreement of the model curves with the \emph{ab initio}
data to be reasonable.
The model explains the transparency induced by the laser in the middle panel of
Fig.~\ref{fig:three-level} in terms of a splitting of the~$1s^{-1} \, 3p$ and
$1s^{-1} \, 3s$~states into an Autler-Townes doublet.
Since the laser causes a suppression of resonant absorption, we call
this electromagnetically induced transparency~(EIT) for
x~rays~\cite{Buth:ET-07}.
Dipole selection rules dictate that the laser can couple only~$1s^{-1} \,
3p_z$ to $1s^{-1} \, 3s$ in a one-photon processes;
so only for~$\vartheta\I{LX} = 0^{\circ}$, EIT occurs.
The suppression for~$\vartheta\I{LX} = 90^{\circ}$ is a consequence of
laser-induced line broadening.

\begin{figure}[bht]
  \begin{center}
    \includegraphics[clip,width=\hsize]{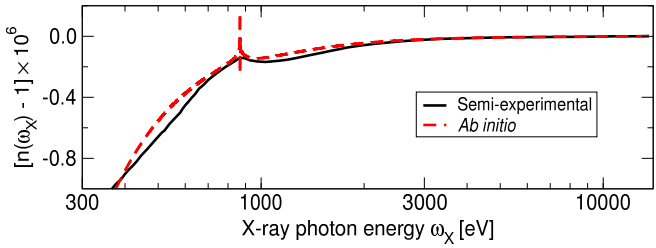}
    \caption{(Color online) Refractive index of neon without laser dressing.}
    \label{fig:NeRI}
  \end{center}
\end{figure}

The refractive index is a classical quantity from Maxwell's equations.
In Fig.~\ref{fig:NeRI}, we show the real part~$n(\omega\I{X})$ for neon
in the x-ray regime without laser dressing at normal temperature and
pressure~\cite{Buth:AR-08}.
It follows from~$n(\omega\I{X}) = 1 + 2 \, \pi \, n_{\#} \,
\alpha\I{I}(\omega\I{X})$, where the number density of neon atoms is~$n_{\#}$.
Further, $\alpha\I{I}(\omega\I{X})$ is the dynamic polarizability which we compute
using~$\hat H\I{I,X} = \alpha \, \hat{\vec p} \mul \vec A\I{X} + \dfrac{1}{2} \, \alpha^2
\, \vec A\I{X}^2$ (in atomic units) for the interaction Hamilton with the x~rays
in dipole approximation~\cite{Buth:AR-08}.
The momentum operator is~$\hat{\vec p}$ and the vector potential is~$\vec A\I{X}$.
We compare our theoretical results with the data from
Fig.~4 of Ref.~\onlinecite{Liggett:RI-68} which was computed from experimental
cross section data.
The agreement is generally good;
only the feature on the $1s \to 3p$~resonance is not reproduced by
the semi-experimental data due to limited resolution.

\begin{figure}
  \begin{center}
    \includegraphics[clip,width=\hsize]{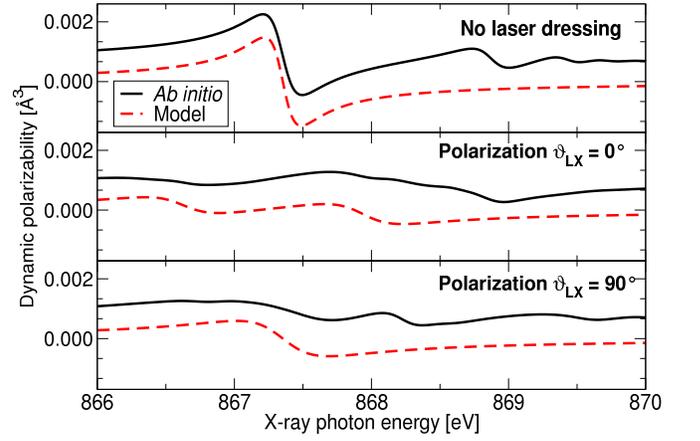}
    \caption{(Color online) Dynamic polarizability of neon near
             the $K$~edge.}
    \label{fig:polarizability}
  \end{center}
\end{figure}

The impact of the laser dressing on neon is investigated for the dynamic
polarizability;
it is plotted in Fig.~\ref{fig:polarizability} for the three cases considered
in Fig.~\ref{fig:NeI_model}.
In this paragraph, we follow Ref.~\onlinecite{Buth:ET-07} and use~$\hat H\I{X}
= \vec r \mul \vec E\I{X}$ in dipole approximation for the interaction with
the x~rays.
The electron position is~$\vec r$ and the electric field of the x~rays
is~$\vec E\I{X}$.
The overall polarizability and its change is small for x~rays.
Laser dressing causes a suppression near the resonance in contrast
to the optical domain.
The $\Lambda$-type model [Fig.~\ref{fig:three-level}] reproduces the structure
of~$\alpha\I{I}(\omega\I{X}, \vartheta\I{LX})$~\cite{Buth:ET-07}.

\begin{figure}[bht]
  \begin{center}
    \includegraphics[clip,width=\hsize]{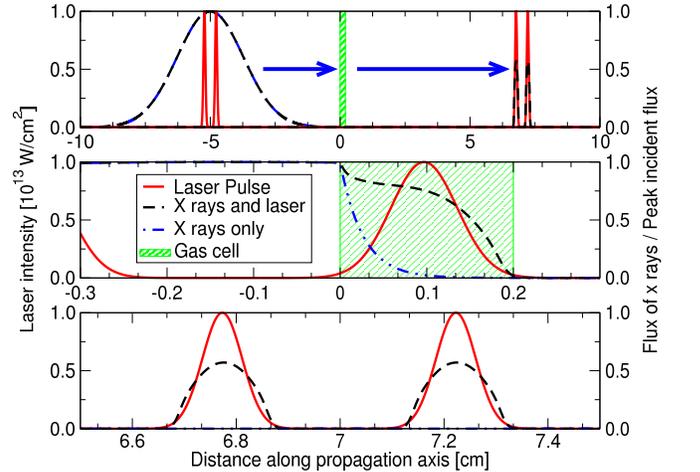}
    \caption{(Color online) EIT-based generation of two ultrashort x-ray pulses
             with well-defined time delay.}
    \label{fig:transmission}
  \end{center}
\end{figure}

EIT for x~rays opens up an exciting novel prospect:
it can be used to imprint pulse shapes of the optical dressing laser onto
a comparatively long x-ray pulse.
Yet only amplitude modulation is practically feasible due to the small
refraction and dispersion in the x-ray domain [Figs.~\ref{fig:NeRI} and
\ref{fig:polarizability}].
In Fig.~\ref{fig:transmission},
the propagation of two-color light through a neon gas cell is shown.
First, the x-ray and laser pulses are still outside the gas cell.
Second, the first of the two laser pulses overlaps with the x-ray pulse
inside the gas cell, thereby substantially enhancing x-ray transmission
in comparison to the laser-off case.
Third, after propagation through the gas cell, two ultrashort
x-ray pulses emerge.
This EIT-based pulse shaping opens up a route to ultrafast all x-ray pump-probe
experiments~\cite{Buth:ET-07}.

\begin{acknowledgments}
C.B.'s research was partly funded by a Feodor Lynen Research Fellowship from
the Alexander von Humboldt Foundation.
Our work was supported by the Office of Basic Energy Sciences,
Office of Science, U.S.~Department of Energy, under Contract
No.~DE-AC02-06CH11357.
\end{acknowledgments}

\end{document}